\documentclass[
reprint,
superscriptaddress,
showpacs,
nofootinbib,
amsmath,amssymb,
aps,
prl,
floatfix,longbibliography
]{revtex4-2}
\usepackage{graphicx}
\usepackage{dcolumn}
\usepackage[hypertexnames, breaklinks=true, bookmarksnumbered=true, bookmarksopen=true, colorlinks=true, linktocpage=true, citecolor=blue, urlcolor=magenta, linkcolor=magenta]{hyperref}
\usepackage{amsmath,amssymb,amsfonts,amsthm}
\usepackage{mathtools} 
\usepackage[T1]{fontenc}
\usepackage[latin9]{inputenc}
\usepackage{txfonts}
\usepackage{bbm} 
\usepackage{bm} 
\usepackage{mathrsfs} 
\usepackage{xcolor}
\usepackage{natbib}
\allowdisplaybreaks

\newcommand{\RR}{\mathbbm{R}}

\DeclareMathOperator{\tr}{Tr}

\newcommand{\id}{\mathbbm{1}}

\theoremstyle{plain}
\newtheorem{theorem}{Theorem}
\newtheorem{lemma}[theorem]{Lemma}

\newcommand{\bra}[1]{\langle #1|}
\newcommand{\ket}[1]{|#1\rangle}
\newcommand{\braket}[2]{\langle #1|#2\rangle}
\newcommand{\ketbra}[2]{|#1\rangle\!\langle#2|}

\newcommand{\I}{\mathscr{I}}
\newcommand{\RS}{\mathscr{R}}
\newcommand{\MI}{\hat{+}}
\newcommand{\MIi}{\hat{-}}

\newcommand{\CM}[1]{{\leavevmode\color{violet}#1}}
\newcommand{\question}[1]{{\leavevmode\color{red}#1}}

\begin{document}


\title{Resource Theory of Imaginarity: New Distributed Scenarios}

\author{Kang-Da Wu}
\affiliation{CAS Key Laboratory of Quantum Information, University of Science and Technology of China, \\ Hefei 230026, People's Republic of China}
\affiliation{CAS Center For Excellence in Quantum Information and Quantum Physics, University of Science and Technology of China, Hefei, 230026, People's Republic of China}

\author{Tulja Varun Kondra}
\affiliation{Centre for Quantum Optical Technologies, Centre of New Technologies, University of Warsaw, Banacha 2c, 02-097 Warsaw, Poland}

\author{Carlo~Maria~Scandolo}
\email{carlomaria.scandolo@ucalgary.ca}
\affiliation{Department of Mathematics and Statistics, University of Calgary, AB, Canada T2N 1N4}
\affiliation{Institute for Quantum Science and Technology, University of Calgary, AB, Canada T2N 1N4}

\author{Swapan Rana}
\affiliation{Physics and Applied Mathematics Unit, Indian Statistical Institute, 203 B T Road, Kolkata 700108, India}

\author{Guo-Yong Xiang}
\email{gyxiang@ustc.edu.cn}
\affiliation{CAS Key Laboratory of Quantum Information, University of Science and Technology of China, \\ Hefei 230026, People's Republic of China}
\affiliation{CAS Center For Excellence in Quantum Information and Quantum Physics, University of Science and Technology of China, Hefei, 230026, People's Republic of China}

\author{Chuan-Feng Li}
\affiliation{CAS Key Laboratory of Quantum Information, University of Science and Technology of China, \\ Hefei 230026, People's Republic of China}
\affiliation{CAS Center For Excellence in Quantum Information and Quantum Physics, University of Science and Technology of China, Hefei, 230026, People's Republic of China}

\author{Guang-Can Guo}
\affiliation{CAS Key Laboratory of Quantum Information, University of Science and Technology of China, \\ Hefei 230026, People's Republic of China}
\affiliation{CAS Center For Excellence in Quantum Information and Quantum Physics, University of Science and Technology of China, Hefei, 230026, People's Republic of China}

\author{Alexander Streltsov}
\email{a.streltsov@cent.uw.edu.pl}
\affiliation{Centre for Quantum Optical Technologies, Centre of New Technologies,
University of Warsaw, Banacha 2c, 02-097 Warsaw, Poland}

\date{\today}

\begin{abstract}  
The resource theory of imaginarity  studies the operational value of imaginary parts in quantum states, operations, and measurements. Here we introduce and study the distillation and conversion of imaginarity in distributed scenario. This arises naturally in bipartite systems where both parties work together to generate the maximum possible imaginarity on one of the subsystems. We give exact solutions to this problem for general qubit states and pure states of arbitrary dimension. We present a scenario that demonstrates the operational advantage of imaginarity: the discrimination of quantum channels without the aid of an ancillary system. We then link this scenario to LOCC discrimination of bipartite states. We experimentally demonstrate the relevant assisted distillation protocol, and show the usefulness of imaginarity in the aforementioned two tasks. 
\end{abstract}


\maketitle

\section{introduction}
Standard quantum theory describes physical reality with complex states, operators, and Hilbert spaces. However, there have always been lots of  questions on the role of complex numbers since the early days of quantum physics \cite{Jordan1993,Stuckelberg, Araki-real,Wootters1990,Entanglement-rebit,Wootters-transport,Plastino1,Plastino2,GisinSimulating,Hardy+Wootters.FoP.2012,Wootters.FoP.2012,Baez,Aleksandrova+2.PRA.2013,Wootters.JPA.2014,Wootters.BC.2016}.  Recently the necessity and usefulness of the imaginary part of quantum mechanics has received significant attention \cite{Hickey+Gour.JPA.2018,BarnumGraydonWilceCCEJA,PRLversion,renou2021quantum,PRAversion,xue2021quantification,Big-experiment,Aberg2006}. Today, quantum mechanics with imaginary numbers seems to be the most successful theory to describe the microscopic world. These research contributions have shown that complex quantum mechanics is fundamentally different from the corresponding real version in many aspects~\cite{renou2021quantum,Stuckelberg,Hardy+Wootters.FoP.2012,Aleksandrova+2.PRA.2013,Wootters.BC.2016,Sperling:21,2021arXiv210805715B,PhysRevA.106.042207,2022arXiv221014443C,PhysRevA.103.L040402}, revealing that the imaginary part is not only necessary for the formulation of quantum theory but also plays an important role in many quantum information tasks~\cite{zhu2020hiding,Wootters.FoP.2012,GisinSimulating}.

The development of quantum information science over the last two decades has led to a reassessment of quantum properties, such as entanglement~\cite{HorodeckiRevModPhys.81.865,PhysRevLett.78.2275} and coherence~\cite{StreltsovRevModPhys.89.041003,BaumgratzPhysRevLett.113.140401}, as resources, which led to the development of quantitative theories that captured these phenomena in a mathematically rigorous fashion \cite{Resource-theories,ChitambarRevModPhys.91.025001}. Nevertheless, imaginarity had not been studied in this framework until the last few years \cite{Hickey+Gour.JPA.2018,PRLversion,PRAversion,xue2021quantification}. In this setting, imaginarity is regarded as a valuable resource that cannot be generated or increased under a restricted class of operations known as \emph{real operations} (RO). Quantum states whose density matrices (in a fixed basis) contain imaginary parts  are viewed as resource states, and thus cannot be created freely by RO. 

In this Letter, we study the resource theory of imaginarity in distributed scenarios. (At least) two parties, Alice (A) and Bob (B) are involved, who share a bipartite state $\rho^{AB}$. In this setting, imaginarity is considered a resource only 
in Bob's system, while Alice can perform arbitrary quantum operations on her system. The duo is further allowed to communicate classically with one another. Overall, we refer to the allowed set of operations in this protocol as \emph{\textbf{L}ocal\textbf{ Q}uantum-\textbf{R}eal operations and \textbf{C}lassical \textbf{C}ommunication (LQRCC)} borrowing the notion from the theory of entanglement~\cite{HorodeckiRevModPhys.81.865} and quantum coherence~\cite{StreltsovRevModPhys.89.041003}. This framework leads to a variety of problems, which we address and solve in this Letter. In particular, we consider assisted imaginarity distillation, where Alice assists Bob in extracting local imaginarity. If only one-way classical communication is used, we provide a solution of this problem for arbitrary two qubit states. We also study assisted state conversion, where the goal is to obtain a specific target state on Bob's side. We solve this problem for any target state, if Alice and Bob share a pure state initially. Furthermore, we study the role of imaginarity in ancilla-free channel discrimination, showing two real channels that are perfectly distinguishable in the ancilla-free scenario once we allow imaginarity, but become  completely indistinguishable if we have access only to real states and real measurements. Additionally, we prove how this task is related to LOCC (Local Operations and Classical Communication) discrimination of quantum states, specifically to the LOCC discrimination of their normalized Choi matrices. Finally, we experimentally implement the above protocols in a quantum photonic setup, performing the proof of principle experiment testing the usefulness of imaginarity in such quantum tasks. Our work opens new avenues towards both theoretical and experimental exploration of imaginarity as a quantum resource.
\medskip{}

\section{resource theory of imaginarity}
The starting point of our work is the resource theory of imaginarity, introduced very recently in Refs.~\cite{Hickey+Gour.JPA.2018,PRLversion,PRAversion}. The free states in imaginarity theory are identified as \emph{real} states, which are real density matrices in a given basis $\left\{\ket{j}\right\}$. The set of all real states is denoted by $\RS$, which can be described by $\RS = \left\{\rho : \braket{j}{\rho|k}\in \RR \textrm{ for all }j,k \right\}$. A quantum operation specified by Kraus operators $\{K_j\}$ satisfying $\sum_jK^\dag_jK_j=\id$, is considered to be free, i.e., \emph{real}, if it contains only real elements in the chosen basis:
$\braket{m}{K_j|n}\in \RR \textrm{ for all }j,m,n$ \cite{Hickey+Gour.JPA.2018,PRLversion}. It is known that the set RO coincides with the set of \emph{completely non-imaginarity creating operations} \cite{Hickey+Gour.JPA.2018}. Moreover, RO coincides with the set of operations which have a \emph{real dilation} \cite{Hickey+Gour.JPA.2018}. The golden unit, i.e.\ the maximally resourceful state, is the same in any Hilbert space, regardless of its dimension. In particular, the maximally imaginary states are the two eigenstates of Pauli matrix $\sigma_y$, \begin{equation}\ket{\hat{\pm}}=\frac{\left(\,\ket{0}\pm i\,\ket{1}\,\right)}{\sqrt{2}}.\end{equation} One maximally imaginary qubit is referred to as an \emph{imbit} in the following. 


Within the framework of quantum resource distillation~\cite{PurificationPhysRevLett.76.722,ChitambarRevModPhys.91.025001,RegulaPhysRevA.101.062315,OneShotCoherenceDistillation}, general quantum states can be used for single-shot or asymptotic distillation of imbits via ROs. In the single-shot regime, the answer was already given in Refs.~\cite{PRLversion,PRAversion}. In particular, the fidelity of imaginarity $F_{\mathrm{I}}$, which quantifies the maximum achievable fidelity between a state $\rho$ and the imbit \begin{equation}F_\mathrm{I}\left(\rho\right)=\max_{\Lambda}F\left(\,\Lambda\left[\rho\right],\ketbra{\MI}{\MI}\,\right),\end{equation} was used as the figure of merit for single-shot distillation, where $F\left(\rho,\sigma\right)=\left[\tr\left(\sqrt{\sigma}\rho\sqrt{\sigma}\right)^{\frac{1}{2}}\right]^2$. The exact value of fidelity of imaginarity for general $\rho$ was shown to be equal to 
\begin{equation}
F_{\mathrm{I}}\left(\rho\right)=\frac{1+\I_R\left(\rho\right)}{2},
\end{equation} where $\I_R\left(\rho\right) =\min_\tau\left\{s \geq 0:\left(\rho+s\tau\right)/\left(1+s\right) \in \RS \right\}$ is the robustness of imaginarity~\cite{PRLversion}. When we consider the asymptotic setting, for large $n$, the fidelity of imaginarity exponentially converges to 1 (for any non-real states). The exponent, for large n, is given by $-\log\left(\tr\sqrt{\rho\rho^{T}}\right) $. For real states, the fidelity of imaginarity is independent of $n$, and is $1/2$ \cite{WuPhysRevLett.121.050401}. Details of the proof can be found in the Appendix.

One of the key motivations for us to study the resource of imaginarity is that we can simulate arbitrary operations or measurements with one imbit at hand, even if all devices allow only real ones in our lab, as we show explicitly in the Appendix. In entanglement theory, one maximally entangled qubit state (ebit) has a clear operational meaning: it can be used to teleport the state of an unknown qubit deterministically to a remote lab. In imaginarity theory, if all the devices are restricted to implement ROs, e.g., we have only half-wave plate in an optical setup~\cite{PRLversion,PRAversion}, we can still prepare arbitrary states or implement arbitrary measurements if we get one imbit at hand. We refer to the Appendix for more details.
\medskip{}
\section{Bipartite imaginarity theory}
The results studied so far concern imaginarity as resource in a single physical system. We now extend our considerations to the bipartite setting. 
As mentioned earlier,  the task involves a bipartite state $\rho^{AB}$ shared by Alice and Bob, and the goal is to maximize imaginarity on Bob's side under LQRCC. If both parties are restricted to real operations, the corresponding set is called local real operations and classical communication (LRCC)~\cite{varun_im}. It is clear that via LQRCC it is possible to create only states of the form 
\begin{equation}
    \rho_\mathrm{qr}=\sum_{j}p_{j}\,\rho_{j}^{A}\otimes\sigma_{j}^{B},
\end{equation}
where $\rho_j^A$ is an arbitrary state on Alice's side, and $\sigma_j^B$ is a real state on Bob's side. States of this form will be called \emph{Quantum-Real (QR)}. In the appendix, we show that the choi matrices corresponding to LQRCC are "invariant" under partial transpose over Bob (Bob is restricted to real operations). This also holds for more general LQRCC maps, which are trace non-increasing (similar to SLOCC in entanglement theory). Using this, we now show that, for arbitrary initial state $\rho_{AB}$ and the target pure state $\ket{\psi_{A'B'}}$, the optimal achievable fidelity for a given probability of success $p$ (given by $F_{p}$), can be upperbounded by a SDP.
\begin{theorem}
Achievable fidelity for a given probablity of success $(F_p(\rho_{AB}\xrightarrow[]{LQRCC}\ket{\psi_{A'B'}})$, of transforming $\rho_{AB}$ into $\ket{\psi_{A'B'}}$ via LQRCC operations can upper bounded by the following semidefinite programme.\\
Maximise:
\begin{equation}
  \frac{1}{p} \tr\left(X_{ABA'B'}\,\rho_{AB}^{T}\otimes \ket{\psi_{A'B'}}\!\bra{\psi_{A'B'}}\right)\,\, 
\end{equation}
under the constraints,
\begin{eqnarray}\nonumber
   && X_{ABA'B'}\geq 0,\, 
     X_{ABA'B'}^{T_{BB'}}=X_{ABA'B'},  \tr_{A'B'} X_{ABA'B'}\leq\id_{AB}\,\,\textrm{and}\,\,\\
     &&\tr\left(X_{ABA'B'}\,\rho_{AB}^{T}\otimes \id_{B'}\right)=p.
\end{eqnarray}
\end{theorem}
In the case of LRCC operations, one has to add an additonal constraint, given by $ X_{ABA'B'}^{T_{AA'}}=X_{ABA'B'}$. For the details about the proof, please refer to the appendix. In the special case when the target state is a local pure state of Bob $\ket{\psi_{B'}}$, one can replace $\ket{\psi_{A'B'}}$ by $\ket{0}\otimes\ket{\psi_{B'}}$, in the objective function. \medskip{} 

\section{Assisted imaginarity distillation}
Having extended the theory of imaginarity to multipartite systems, we are now ready to present assisted imaginarity distillation. In this task, Alice and Bob aim to extract imaginarity on Bob's side by applying LQRCC operations, which is in analogy to assisted entanglement distillation~\cite{DiVincenzo1999,SmolinPhysRevA.72.052317,PhysRevA.73.062331} and assisted distillation of quantum coherence~\cite{AssistedPhysRevLett.116.070402}. We assume that Alice and Bob share an arbitrary mixed state $\rho^{AB}$, and the process is performed on a single copy of the state and only one-way classical communication from Alice to Bob is used. If Alice performs a general measurement $\left\{M_j^A\right\}$ on her side, the probability $p_j$ and the corresponding post-measurement state of Bob $\rho_j^B$ are given respectively by $p_{j}=\tr\left[\left(M_{j}^{A}\otimes\id^B\right)\rho^{AB}\right]$, $\rho_{j}^{B} =1/p_{j}\tr_{A}\left[\left(M_{j}^{A}\otimes\id^B\right)\rho^{AB}\right]$.

As a figure of merit we now introduce the \emph{assisted fidelity of imaginarity}, quantifying the maximal single-shot fidelity between Bob's final state and the maximally imaginary state $\ket{\hat{+}}$: \begin{equation}F_{\mathrm{a}}\left(\rho^{AB}\right)=\max_{\left\{M_j^A,\,\Lambda_j\right\}}\,\sum_{j}p_{j}F\left(\,\Lambda_{j}\left[\,\rho_{j}^{B}\,\right],\,\ketbra{\MI}{\MI}\,\right).\end{equation} The maximum is taken over all POVMs on Alice's side, and all real operations $\Lambda_j$ on Bob's side. 
For two-qubit states, we can derive the exact analytic expression. Consider a two-qubit state $\rho^{AB}$, which can be written as
$\rho=\left(\id_4+\boldsymbol{a}\cdot\boldsymbol{\sigma}\otimes \id+\id\otimes \boldsymbol{b}\cdot\boldsymbol{\sigma}+\sum_{k,l}E_{kl}\sigma_{k}\otimes\sigma_l\right)/4$, 
where the $\sigma_k$'s are Pauli matrices, $\boldsymbol{a}=\left(a_1,a_2,a_3\right)$ and $\boldsymbol{b}=\left(b_1,b_2,b_3\right)$ describe local Bloch vectors of Alice and Bob, respectively, and $E_{kl}=\tr \left(\sigma_k\otimes\sigma_l\rho\right)$. Equipped with these tools, we are now ready to give a closed expression for the assisted fidelity of imaginarity for all two-qubit states. 

\begin{theorem} \label{thm:FaQubit}
For any two-qubit state $\rho^{AB}$ the assisted fidelity of imaginarity is given by
\begin{equation}
F_{\mathrm{a}}\left(\rho^{AB}\right)=\dfrac{1}{2}\left(1+\max\left\{\left|b_2\right|,\left|\boldsymbol{s}\right|\right\}\right). \label{eq:FaQubits}
\end{equation}
where the vector $\boldsymbol{s}=\left(E_{12},E_{22},E_{32}\right)$.
\end{theorem}
The proof is presented in the Appendix.   

We will now extend our results to stochastic state transformations, where the goal is to achieve a transformation with the maximum possible probability. 
To this end, we introduce the \textit{geometric measure of imaginarity} and the \textit{concurrence of imaginarity}, presented  in Refs.~\cite{Uhlmann_2000,varun_im} respectively as
\begin{subequations}\begin{align}
&\mathscr{I}_g\left(\rho\right)=\frac{1-\sqrt{F\left(\rho,\rho^T\right)}}{2},\\
&\mathscr{I}_c\left(\rho\right)=\max\left\{0,\lambda_1-\sum_{j>1}\lambda_j\right\},\end{align}\end{subequations} where $\left\{\lambda_1,\lambda_2,\dots\right\}$ are the eigenvalues (in decreasing order) of $\left(\sqrt{\rho}\rho^T\sqrt{\rho}\right)^{\frac{1}{2}}$. With this in place, we now extend this scenario to the bipartite regime where we will show how Alice can assist Bob ($\rho^B$) to get the target state $\sigma^B$ with optimal probability. Now we use the following parameterization: $\sin^2\alpha=\left[1-\mathscr{I}_c\left(\rho^B\right)\right]/2$ and $\sin^2\beta=\mathscr{I}_g\left(\sigma^B\right)$ with $\alpha,\,\beta\in(0,\frac{\pi}{2})$.

\begin{lemma}\label{Theo:optimalprobability}
 For any bipartite pure state $\psi^{AB}$, the optimal probability of Bob preparing a local state $\sigma^B$, getting assistance from Alice, is given by
\begin{equation}
    P\left(\psi^{AB}\rightarrow\sigma^B\right)=\min\left\{\dfrac{\sin^2\alpha}{\sin^2\beta},1\right\}.
\end{equation}
\end{lemma}
The proof of Lemma~\ref{Theo:optimalprobability} is presented in the Appendix. In Ref.~\cite{varun_im} the authors provided tight continuity bounds for the geometric measure. Using these bounds, along with Lemma~\ref{Theo:optimalprobability}, we can provide an analytical expression for the optimal probability of Bob preparing a local state with an allowed error, with assistance from Alice. 
Similarly, we can also find a closed expression for the optimal achievable fidelity, for a given probability of success. The following theorem collects these results.
\begin{theorem} \label{thm:PureConversion}
For any bipartite pure state $\psi^{AB}$, the optimal probability $P_{f}$ of Bob preparing a local state $\sigma^{B}$, with a fidelity $f$ via assistance from Alice, is given by
\begin{equation}\label{eq:optimal_probability_imaginarity}
P_{f}\left(\,\psi^{AB}\rightarrow\sigma^{B}\,\right)=\begin{cases}
1 & \textrm{for }\alpha-\beta +\gamma\geq0\\
\dfrac{\sin^2\alpha}{\sin^{2}\left(\,\beta-\gamma\,\right)} & \textrm{otherwise}\end{cases}
\end{equation}
where $\gamma=\cos^{-1}\sqrt{f}$.

The optimal achievable fidelity for a given probability of success $p$, can be expressed as:
\begin{equation}
F_{p}\left(\,\psi^{AB}\rightarrow\sigma^B\,\right)=\begin{cases}
1\,\,\,&\mathrm{for}\,\,\,p\leq\dfrac{\sin^2\alpha}{\sin^2\beta}\\

\cos^{2}\left[\beta-\sin^{-1}\!\left(\dfrac{\sin\alpha}{\sqrt{p}}\right)\right]\,\, &\mathrm{otherwise}.\label{eq:optimal_fidelity_imaginarity}
\end{cases}
\end{equation}
\end{theorem}
Details of the proof for the above theorem can be found in the Appendix.

\medskip{}

\textit{Imaginarity in channel discrimination}---We will now discuss the role of imaginarity in channel discrimination. Specifically, here we focus on the variant of channel discrimination which we call \emph{ancilla-free}, in that it does not involve an ancillary system (cf.\ Refs.~\cite{TakagiPhysRevLett.122.140402,Takagi+Regula.PRX.2019}). It can be regarded as a game, where one has access to a ``black box'' with the promise that it implements a quantum channel $\Lambda_j$ with probability $p_j$. The goal of the game is to guess $\Lambda_j$ by choosing optimal initial state $\rho$ and positive operator-valued measure (POVM) $\left\{M_j\right\}$, which is used to distinguish the $\Lambda_j\left(\rho\right)$'s. Theoretically, the probability of guessing the channel $\Lambda_j$ correctly is given as 
\begin{equation}
p_{\mathrm{succ}}\left(\rho,\left\{p_j,\Lambda_j\right\},\left\{M_j\right\}\right)=\sum_j p_j \tr \left[M_j\Lambda_j\left(\rho\right)\right].    
\end{equation}
Recently, it has been shown that \emph{any} quantum resource has an operational advantage in the channel discrimination task~\cite{Takagi+Regula.PRX.2019,TakagiPhysRevLett.122.140402}, namely a resource state $\rho$ (i.e.\ a quantum state that is not free) outperforms any free $\sigma$ in a specific channel discrimination task.

Now we put the above protocol into imaginarity theory by considering the task of discrimination of real channels. To see an advantage, we need imaginarity both in the probe state and in the measurement, since, as we show in the Appendix, this task is equivalent to LOCC discrimination of their corresponding normalized Choi states, in which we need imaginarity in the measurements of both particles. To better illustrate this idea, we will provide an example of two real channels that cannot be distinguished in the ancilla-free scenario by using only real states and measurements, but they become instead perfectly distinguishable once we have access to imaginarity for states and measurements. To this end, let us consider two real qubit channels prepared with equal probability: \begin{equation}\begin{aligned}\mathcal{N}\,: \,& \rho\mapsto\frac{1}{2}\left(\,\rho+\sigma_x\,\sigma_z\,\rho\,\sigma_z\,\sigma_x\,\right),\\
\mathcal{M}\,:\, & \rho\mapsto\frac{1}{2}\left(\,\sigma_x\,\rho\,\sigma_x+\sigma_z\,\rho\,\sigma_z\,\right),
\end{aligned}
\label{eq:ChannelDiscriminationChannels}\end{equation}where $\sigma_x$ and $\sigma_z$ are Pauli matrices.
If we input a real state $\rho$ into either of these two channels, they will produce exactly the same output $\id/2$, thus we cannot distinguish them better than making a random guess, even if we allowed imaginarity in our measurements. On the other hand, if imaginarity is forbidden in measurements, no matter how we choose the probe state (even if itis non-real), we cannot still distinguish them at all, because the only way to discriminate between the outputs of the two channels would be to perform a measurement associated with the $\sigma_y$ Pauli matrix. Indeed, if the  probe state has an off-diagonal entry $\rho_{01}$ with non-zero imaginary part, wherever the output of $\mathcal{N}$ has $\mathrm{Im}\,\rho_{01}$, the output of $\mathcal{M}$ will show $-\mathrm{Im}\,\rho_{01}$ in its place. 
Only if we implement a projective measurement of $\sigma_y$ can we perfectly distinguish these two channels. Therefore, the only way to achieve a success probability better than random guessing is to introduce imaginarity into both the initial state $\rho$ and the measurement. 

It is worth noting that the same two channels $\mathcal{N}$ and $\mathcal{M}$ become perfectly distinguishable even with no imaginarity in the probe state and in the measurement if we remove the requirement of ancilla-free discrimination. If we allow an ancilla $R$, we need to consider a bipartite input state $\rho^{RA}$ and a bipartite POVM $\left\{M_1^{RA},M_2^{RA}\right\}$, with success probability 
\begin{equation}
p_{\mathrm{succ}}\left(\rho,\left\{\frac{1}{2},\Lambda_j\right\},\left\{M_j\right\}\right)=\frac{1}{2}\sum_{j=1}^2 \tr \left[M_j^{RA}\left(\mathcal{I}^{R}\otimes\Lambda_j\right)\left(\rho^{RA}\right)\right], 
\end{equation}
where $\Lambda_1=\mathcal{N}$ and $\Lambda_2=\mathcal{M}$. Now, let us take $\rho^{RA}=\phi^+=\ketbra{\phi^+}{\phi^+}$, with  $\ket{\phi^+}=\frac{1}{\sqrt{2}}\left(\ket{00}+\ket{11}\right)$. If we feed $\phi^+$ to both channels, we get
\begin{equation}\label{eqs:Choistate1}
\begin{aligned}
\mathcal{I}\otimes\mathcal{N}\left(\phi^+\right)& =\frac{1}{2}\left(\ket{\phi^+}\!\bra{\phi^+} + \ket{\psi^-}\!\bra{\psi^-}\right),\\\mathcal{I}\otimes\mathcal{M}\left(\phi^+\right)& =\frac{1}{2}\left(\ket{\phi^-}\!\bra{\phi^-}+\ket{\psi^+}\!\bra{\psi^+}\right),
\end{aligned}
\end{equation}
where $\ket{\phi^-}=\frac{1}{\sqrt{2}}\left(\ket{00}-\ket{11}\right)$, $\ket{\psi^+}=\frac{1}{\sqrt{2}}\left(\ket{01}+\ket{10}\right)$, and $\ket{\psi^-}=\frac{1}{\sqrt{2}}\left(\ket{01}-\ket{10}\right)$. As noted in Ref.~\cite{PRLversion}, these two output states can be perfectly distinguished by the real POVM $\left\{M_1,M_2\right\}$, where
\begin{equation}\label{eqs:optimalM}
\begin{aligned}
    M_1 &= \ketbra{\MI}{\MI}\otimes \ketbra{\MIi}{\MIi} + \ketbra{\MIi}{\MIi}\otimes \ketbra{\MI}{\MI}, \\
    M_2 &= \ketbra{\MI}{\MI}\otimes \ketbra{\MI}{\MI} + \ketbra{\MIi}{\MIi}\otimes \ketbra{\MIi}{\MIi}.
\end{aligned}
\end{equation}
This shows that the two real channels can be distinguished perfectly with the aid of an ancilla, only using real states and real measurements.

\begin{figure}
	\centering
	\includegraphics[scale=0.08]{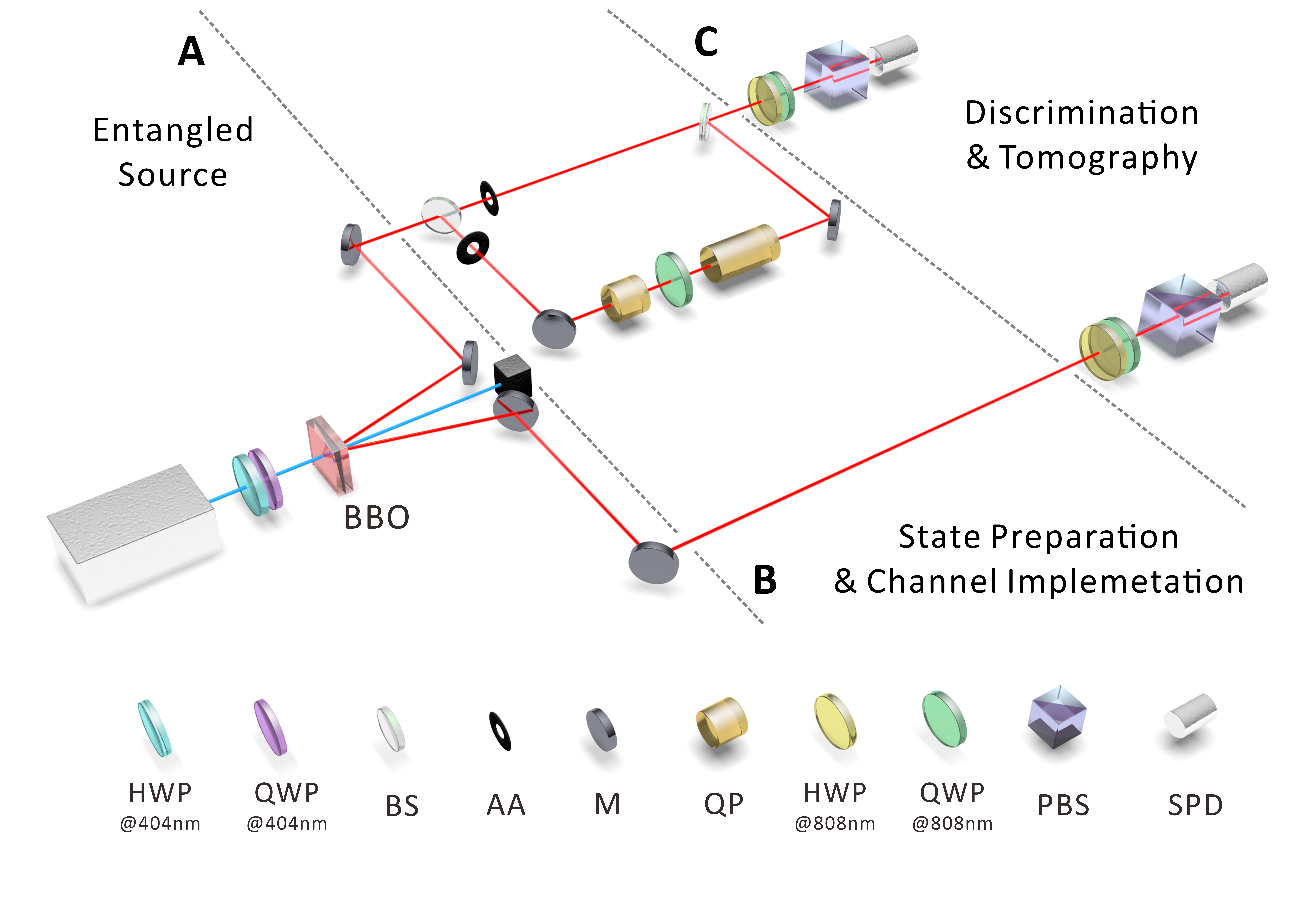}
	\caption{\label{fig:exp}
		\textbf{Experimental setup}. The whole experimental setup is divided into three modules: \textbf{A} Entangled source, \textbf{B} state preparation \& channel implementation, and \textbf{C}  discrimination \& tomography. The optical components include: QP, quartz plate; SPD, single photon detectors; BS, beamsplitters; AA, adjustable aperture; PBS, polarizing beamsplitter; QWP, quarter-wave plate; HWP, half-wave plate.
	}
\end{figure}
\medskip{}

\section{Experiments}
We experimentally implement the aforementioned assisted imaginarity distillation and channel discrimination protocols. The whole experimental setup is illustrated in Fig.~\ref{fig:exp}, which consists of three modules: module \textbf{A} enables us to prepare a two-qubit entangled state via spontaneous parametric down conversion (SPDC) process:
\begin{equation}\label{eq:Purestate}
\ket{\psi}^{AB}=a\,\ket{00}+b\,\ket{11},
\end{equation}
with arbitrary $a$ and $b$ with $\left|a\right|^2+\left|b\right|^2=1$ which can be tuned by changing the angles of 404 nm HWP and QWP. Note that we have conventionally set $\ket{0}:=\ket{H}$ and $\ket{1}:=\ket{V}$. Module \textbf{B} utilizes an unbalanced Mach-Zehnder interferometer together with module \textbf{A} to prepare a class of Werner states:
\begin{equation}\label{eq:Wernerstate}
\rho^{AB}=p\,\ketbra{\phi^+}{\phi^+}+\left(1-p\right)\dfrac{\id}{4},
\end{equation}
where $p$ denotes the purity of the two-qubit state. Module \textbf{B} also allow us to implement single-qubit channels in ancilla-free scenario. Module \textbf{C} allows us to perform quantum-state tomography (QST) to identify the final two-qubit polarization-encoded states concerned, or perform assisted imaginarity distillation by performing local measurement on Alice's photons and identifying the exact amount of imaginarity by QST of Bob's state. Moreover, this module allows us to implement channel discrimination by performing local measurement on the polarization state of a single-photon when the other is used as a trigger. We refer to the Appendix for more details.

\begin{figure}
	\centering
	\includegraphics[scale=0.28]{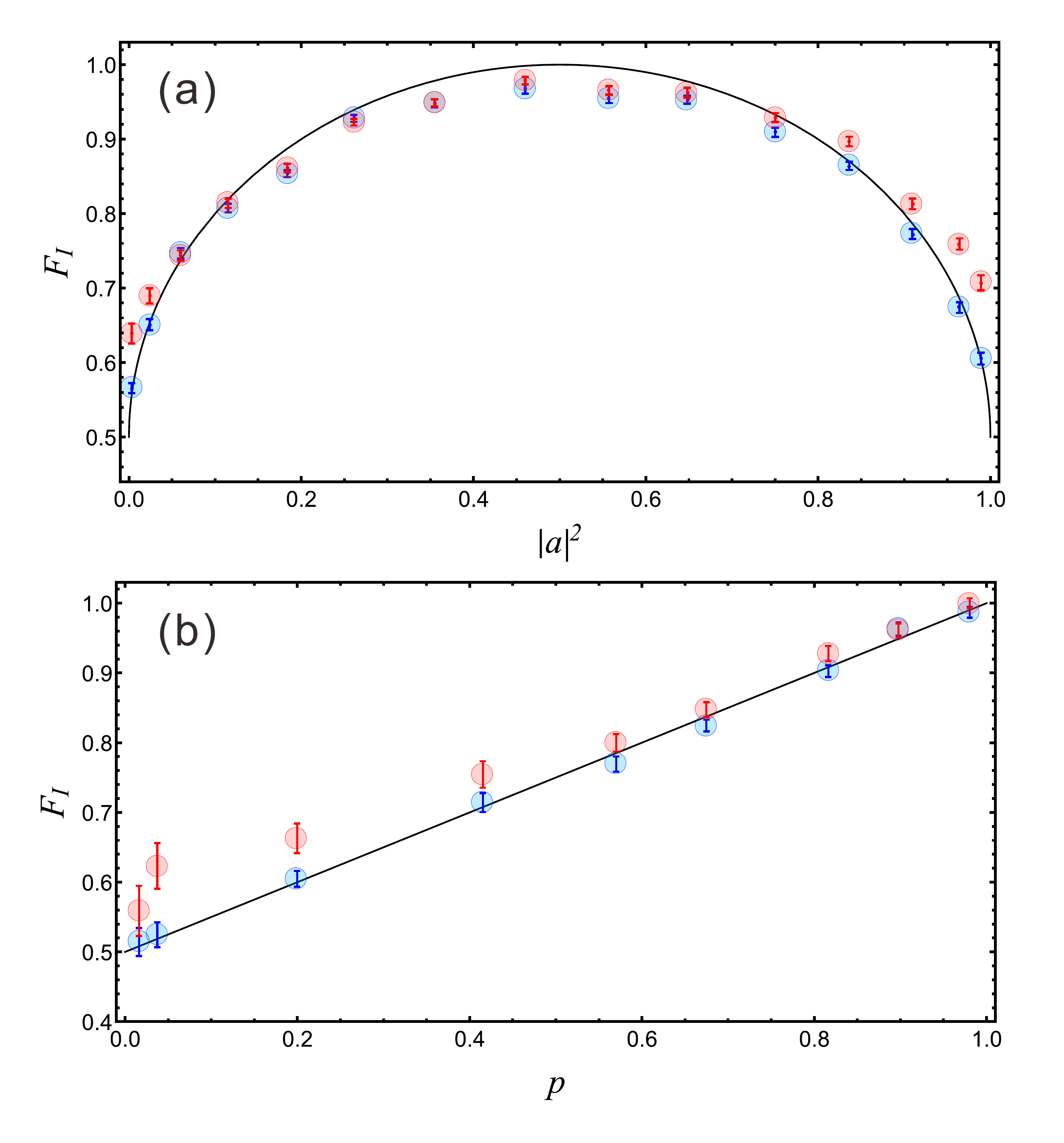}
	\caption{\label{fig:assisted}
		\textbf{Experimental results for assisted imaginarity distillation}. (a) Initial pure states $\ket{\psi}^{AB}=a\ket{00}+b\ket{11}$; (b) initial Werner states $\rho^{AB}=p\ketbra{\phi^+}{\phi^+}+\left(1-p\right)\id/4$. In both experiments, red disks represent the calculated fidelity of imaginarity by assistance using Theorem~\ref{thm:FaQubit}  for experimentally reconstructed two-qubit states, and blue disks represent actual obtained average fidelity of imaginarity in experiments using the optimal measurement on Alice's system.
	}
\end{figure}

\begin{figure}
	\centering
	\includegraphics[scale=0.28]{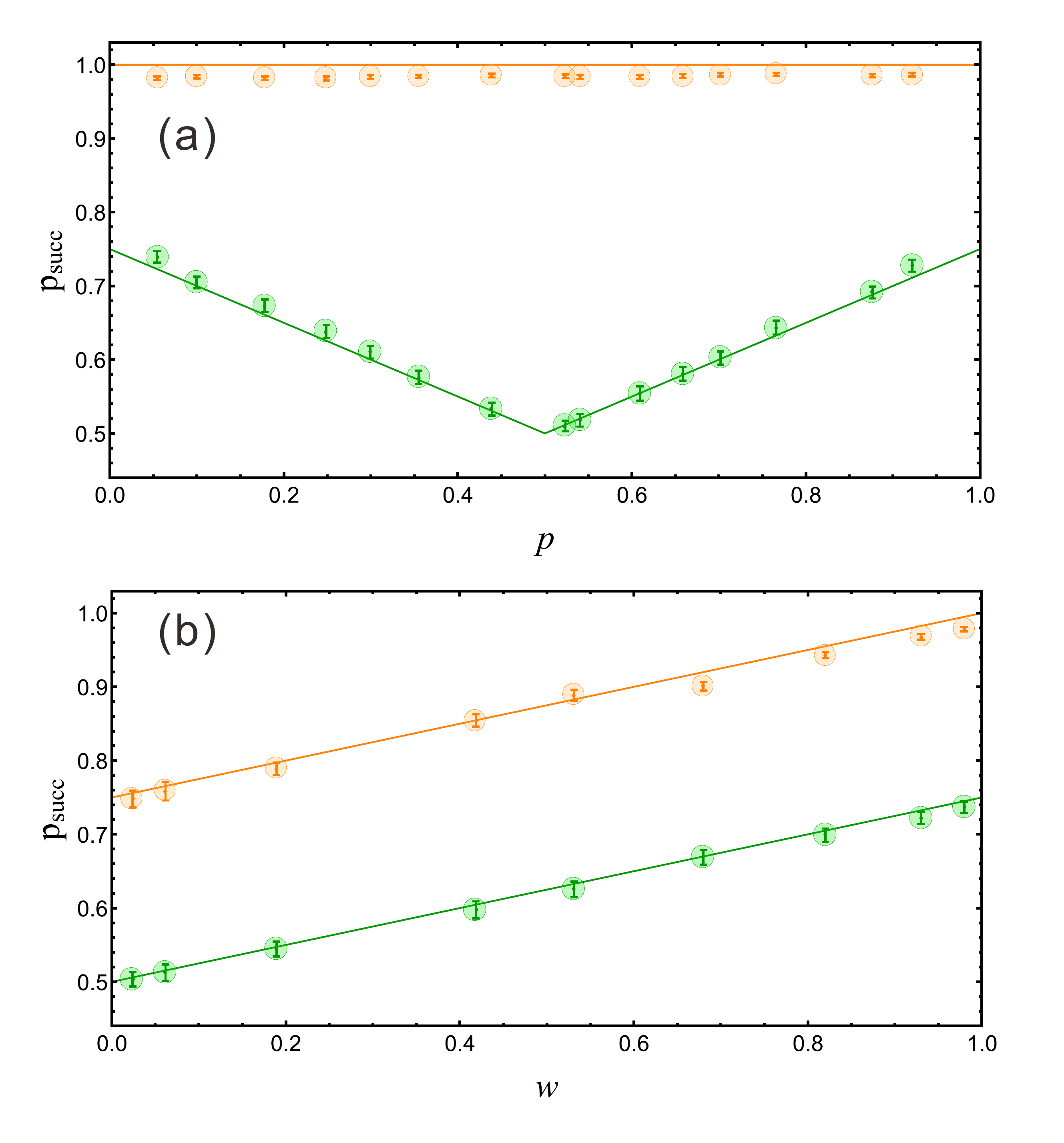}
	\caption{\label{fig:discrimination}
		\textbf{Experimental results for discrimination tasks}. Two channel discrimination tasks are tested : (a) $\mathcal{M}_p\left(\rho\right)=p\rho+\left(1-p\right)\sigma_x\sigma_z\rho\sigma_z\sigma_x$,  $\mathcal{N}\left(\rho\right)=\left(\sigma_x\rho\sigma_x+\sigma_z\rho\sigma_z\right)/2$. Using imaginarity one can perfectly distinguish the two channels. However, if only real operators are allowed, then the optimal guessing probability is $1/2+\left|2p-1\right|/2$; (b) $\mathcal{M}_w\left(\rho\right)=w\rho+\left(1-w\right)\id/2$, 
 $\mathcal{N}\left(\rho\right)=\left(\sigma_x\rho\sigma_x+\sigma_z\rho\sigma_z\right)/2$. The optimal probabilities for successful guessing are $3/4+w/4$ and $1/4+w/4$ for the case where imaginarity is allowed, and where only real states and measurements are allowed, respectively.
	}
\end{figure}

We then perform proof of principle experiments of the one-shot assisted imaginarity distillation and the ancilla-free channel discrimination tasks. Results are shown in Figs.~\ref{fig:assisted} and \ref{fig:discrimination} respectively. 

For assisted imaginarity distillation, we experimentally prepare two classes of two-qubit states. The first class of states as in Eq.~(\ref{eq:Purestate}). Theoretically, the upper bound for single-shot assisted imaginarity distillation can be calculated from Theorem~\ref{thm:FaQubit} as $F_{\mathrm{I}}\left(\ket{\psi}^{AB}\right)=2\left|ab\right|$. From Fig.~\ref{fig:assisted}(a), we can see that the experimentally obtained average imaginarity after assistance (blue disks) approximately equals to the experimentally obtained upper bound (red disks) within reasonable experimental imperfections. The second class of states are generated as Werner states in Eq.~(\ref{eq:Wernerstate}). Theoretically, the maximum average fidelity of imaginarity after assistance is calculated as $F_{\mathrm{I}}(\rho^{AB})=p$. Fig.~\ref{fig:assisted}(b) details the relevant experimental results. From both results we see that experimentally obtained average fidelity of imaginarity data and upper bound obtained from two-qubit state tomography agree well with theoretical predictions.  

We then show the usefulness of imaginarity in channel discrimination for various discrimination tasks. Fig.~\ref{fig:discrimination}
details these results for two discrimination tasks. The first discrimination task involves  two channels  given by
\begin{equation}
\begin{aligned}\label{eq:Channeltask1}
 &\mathcal{M}\left(\,\rho,\,p\,\right)=p\rho+\left(1-p\right)\sigma_x\,\sigma_z\,\rho\,\sigma_z\,\sigma_x,\\
 &\mathcal{N}\left(\rho\right)=\frac{1}{2}\left(\sigma_x\,\rho\,\sigma_x+\sigma_z\,\rho\,\sigma_z\right).
\end{aligned}\end{equation}
Note that the two channels preserve real density matrices. The experimental results of this discrimination task are shown in Fig.~\ref{fig:discrimination}(a). If we can use imaginarity in measurements and initial states, we can perfectly distinguish the two channels [orange disks in Fig.~\ref{fig:discrimination}(a)]. However, if we allow only real density matrices as initial states or real measurement operators, we get a theoretical optimal guessing probability of $1/2+\left|2p-1\right|/4$ for the ancilla-free channel discrimination. Experimental data are in agreement with the theoretical predictions [see green disks in Fig.\ref{fig:discrimination}(a)]. Here we note that the two channels are exactly the same as in Eqs.~\eqref{eq:ChannelDiscriminationChannels} when $p=1/2$.

For the second discrimination task, we consider
\begin{equation}
\begin{aligned}
    &\mathcal{M}\left(\,\rho,\,w\,\right)=w\,\rho+\left(1-w\right)\dfrac{\id}{2},\\
 &\mathcal{N}\left(\rho\right)=\frac{1}{2}\left(\sigma_x\,\rho\,\sigma_x+\sigma_z\,\rho\,\sigma_z\right).
\end{aligned}\end{equation}
The results are shown in Fig.~\ref{fig:discrimination}(b). If non-real states and measurement operators are allowed, then we get a theoretical optimal distinguishing probability as $3/4+w/4$, which is plotted as the upper orange line in Fig.~\ref{fig:discrimination}(b). The relevant experimentally obtained distinguishing probabilities are shown as orange disks. If imaginarity is prohibited in this task, then the optimal distinguishing probability reads $1/2+w/4$, and is plotted as the lower green line, together with experimental values represented by green disks. We can draw a similar conclusion to the first discrimination task.

\medskip{}

\section{discussion}
The results presented above are mainly based on the new set of LQRCC operations which was introduced and studied in this article. We considered assisted imaginarity distillation in this setting, and completely solved the problem for general two-qubit states. Moreover, we discussed the task of single-shot assisted imaginarity distillation for arbitrary pure states in higher dimensions. The usefulness of imaginarity in channel discrimination is both theoretically and experimentally shown for a class of real channels. 

There are in fact many scenarios of practical relevance where the task of assisted imaginarity distillation can play a central role. For instance, think of a remote or unaccessible system on which imaginarity is needed as a resource (e.g., in the task of local discrimination of quantum states): our results give optimal prescriptions to inject such imaginarity on the remote target by acting on an ancilla. The results provide insight into both the operational characterization as well as the mathematical formalism of the resource theory of imaginarity, contributing to a better
understanding of this fundamental resource.

\medskip{}
\begin{acknowledgments}
 The work at the University of Science and Technology of China is supported by the National Key Research and Development Program of China (No. 2018YFA0306400), the National Natural Science Foundation of China (Grants Nos. 12134014, 12104439, 61905234, 11974335, 11574291, and 11774334), the Key Research Program of Frontier Sciences, CAS (Grant No. QYZDYSSW-SLH003), USTC Research Funds of the Double First-Class Initiative (Grant No. YD2030002007) and the Fundamental Research Funds for the Central Universities (Grant No. WK2470000035, WK2030000063). The work at Poland was supported by the National Science Centre, Poland, within the QuantERA II Programme (No 2021/03/Y/ST2/00178, acronym ExTRaQT) that has received funding from the European Union's Horizon 2020 research and innovation programme under Grant Agreement No 101017733 and the ``Quantum Optical Technologies'' project, carried out within the International Research Agendas programme of the Foundation for Polish Science co-financed by the European Union under the European Regional Development Fund. CMS acknowledges the support of the Natural Sciences and Engineering Research Council of Canada (NSERC) through the Discovery Grant ``The power of quantum resources'' RGPIN-2022-03025 and the Discovery Launch Supplement DGECR-2022-00119. 
\end{acknowledgments}

\appendix

\section{Implementing general quantum operations}
Here, we show that one imbit is necessary and sufficient to implement arbitrary quantum operation. To see this, let's say we want to implement a quantum operation $\Lambda$ on $\rho$ with Kraus operators given by $\{K_j\}$, such that $\sum_jK_j^{\dagger}K_j=P\leq \id$. To implement this, we construct a real quantum operation ($\Lambda_r$) with Kraus operators given by $\{K_{j}\otimes\ket{\hat{+}}\bra{\hat{+}}+K^{*}_{j}\otimes\ket{\hat{-}}\bra{\hat{-}}\}$. It is easy to see that
\begin{equation}
    \Lambda_r(\rho\otimes\ket{\hat{+}}\bra{\hat{+}})=\Lambda(\rho)\otimes\ket{\hat{+}}\bra{\hat{+}}
\end{equation}
and
\begin{eqnarray}
   \sum_{j}(K^{\dagger}_{j}\otimes\ket{\hat{+}}\bra{\hat{+}}+K^{T}_{j}\otimes\ket{\hat{-}}\bra{\hat{-}})(K_{j}\otimes\ket{\hat{+}}\bra{\hat{+}}+K^{*}_{j}\otimes\ket{\hat{-}}\bra{\hat{-}})\nonumber\\=P\otimes\ket{\hat{+}}\bra{\hat{+}}+P^{T}\otimes\ket{\hat{-}}\bra{\hat{-}}\leq\id\otimes\id\nonumber
\end{eqnarray}
The last inequality follows from the fact that,
\begin{equation}
    P\leq \id \iff P^{T}\leq \id.
\end{equation}
 This shows that one imbit is sufficient to implement general quantum operations. Now we show that, there exists a quantum channel, which necessarily requires one imbit, to implement via real operations. As an example, consider the following map ($\Lambda_+$) given by
 \begin{equation}
     \Lambda_+(\rho)=\ket{\hat{+}}\bra{\hat{+}}\,\,\textrm{forall}\,\,\rho.
 \end{equation}
We now show, by contradiction, that the above quantum map requires one imbit to implement. Let's say there is a implementation (with a real operation $\Lambda'_r$) such that,
\begin{equation}\label{contradiction}
    \Lambda'_r(\rho\otimes\sigma)= \Lambda_+(\rho)=\ket{\hat{+}}\bra{\hat{+}}
\end{equation}
here, if $\sigma$ is not an imbit and $\rho=\ket{0}\bra{0}$, its easy to see that the state transformation in Eq. (\ref{contradiction}) is not possible. This is because $\mathscr{I}_g(\ket{0}\bra{0}\otimes\sigma)=\mathscr{I}_g(\sigma)<\mathscr{I}_g(\ket{\hat{+}}\bra{\hat{+}})$.

\section{Properties of LRCC operations}
For any real CP map $\Lambda:R\rightarrow R'$, $\Gamma^{\Lambda}_{RR'}$ is the corresponding choi matrix of $\Gamma^\Lambda_{RR'}$, given by
\begin{equation}
   \Gamma^\Lambda_{RR'}=\openone\otimes\Lambda\left(\sum_{j,k}\ket{j}\!\bra{k}\otimes\ket{j}\!\bra{k}\right).
\end{equation}
Any LQRCC map ($\Lambda$) can be represented in the following way
\begin{eqnarray}\label{LQRCC}
    \Lambda = \sum_i \Lambda_i\otimes\Lambda^r_i.
\end{eqnarray}
Here, $\lambda_i$ is a CP (trace non increasing) map acting locally on Alice's hilbert spapce and $\Lambda^r_i$ is a local real CP map on Bob's hilbert space. The choi matrix of $\Lambda_i\otimes\Lambda^r_i$ is given by
\begin{eqnarray}
    \Gamma^{\Lambda_i\otimes\Lambda^r_i}_{AB\rightarrow A'B'} &= \openone_{AB}\otimes\Lambda_i\otimes\Lambda^r_i\left(\sum_{j,j',k,k'}\ket{jk}\!\bra{j'k'}\otimes\ket{jk}\!\bra{j'k'}\right)\nonumber\\ \nonumber
    &=\sum_{j,j',k,k'}\ket{j}\bra{j'}\otimes\ket{k}\bra{k'}\otimes\Lambda_i(\ket{j}\bra{j'})\otimes\Lambda^r_i(\ket{k}\bra{k'})
\end{eqnarray}
Let's now take the transpose of this choi matrix over $BB'$
\begin{eqnarray}
    &(\Gamma^{\Lambda_i\otimes\Lambda^r_i}_{AB\rightarrow A'B'})^{T_{BB'}}=\sum_{j,j',k,k'}\ket{jk'}\bra{j'k}\otimes\Lambda_i(\ket{j}\bra{j'})\otimes(\Lambda^r_i(\ket{k}\bra{k'}))^T\nonumber\\
   & =\sum_{j,j',k,k'}\ket{jk'}\bra{j'k}\otimes\Lambda_i(\ket{j}\bra{j'})\otimes\Lambda^r_i(\ket{k'}\bra{k})\nonumber\\
   &= \Gamma^{\Lambda_i\otimes\Lambda^r_i}_{AB\rightarrow A'B'}
\end{eqnarray}
In the second line we used the fact that, real operations commute with transpose. Since any LQRCC operation can be represented as (\ref{LQRCC}), the choi matric of any LQRCC operation is invariant under partial transpose over Bob's systems. For LRCC operations, additionally the choi matrix is always real.
\section{Proof of Theorem 1}
In the following, we assume that $A$ and $B$ is a qubit. A general two-qubit state $\rho^{AB}$ can be written as
\begin{equation}\label{2qubit}
\rho=\frac{1}{4}\left(\id\!\otimes\! \id+\sum_{k}a_k\sigma_k\!\otimes \!\id+\sum_{l}b_l \id\!\otimes\! \sigma_l+\sum_{k,l}E_{kl}\sigma_k\!\otimes\!\sigma_l\right),
\end{equation}
where $\boldsymbol{a}=(a_1,a_2,a_3)$ and $\boldsymbol{b}=(b_1,b_2,b_3)$ are local Bloch vectors of Alice and Bob, respectively, and $E_{kl}=\tr (\sigma_k\otimes\sigma_l\rho)$. A general single-qubit POVM element on Alice's side can be written as 
\begin{equation}
    M_n^A=q_n\left(\id+\sum_j\alpha_{nj}\sigma_j\right) \label{eq:QubitPOVM}
\end{equation}
with probabilities $0\leq q_n\leq 1$, $\sum_n q_n = 1$, and vectors $\boldsymbol{\alpha}_n$ such that $|\boldsymbol{\alpha}_n|\leq 1$ and $\sum_n q_n\boldsymbol{\alpha}_n=0$. The  measurement $\{M_n^A\}$ gives outcome $n$ with probability 
\begin{equation}
p_n=q_n\left(1+\boldsymbol{a}\cdot\boldsymbol{\alpha}_n\right), \label{eq:pi}
\end{equation}
and the Bloch vector of Bob's post-measurement state is
\begin{equation}
\boldsymbol{b}_n=\frac{\boldsymbol{b}+ E^T\boldsymbol{\alpha}_n}{1+\boldsymbol{a}\cdot\boldsymbol{\alpha}_n}. \label{eq:bi}
\end{equation}

After Alice communicates her measurement outcome $n$ to Bob, he applies a real operation $\Lambda_n$ to his post-measurement state $\rho_i^B$. For each measurement outcome $n$, Bob aims to maximize the fidelity between $\Lambda_n[\rho_n^B]$ and the maximally imaginary state $\ket{\hat{+}}$. The maximum is given by the fidelity of imaginarity $F_\mathrm{I}$ which for single-qubit states $\rho_n^B$ reduces to
\begin{equation}
    F_\mathrm I (\rho_n^B)=\frac{1}{2}\left(1+\left|\tr[\rho_n^B\sigma_2]\right|\right).
\end{equation}
Using this result together with Eqs.~(\ref{eq:pi}) and (\ref{eq:bi}) we can express our figure of merit $F_{\mathrm{a}}$ as follows:
\begin{align}
F_{\mathrm{a}} (\rho^{AB})&=\max_{M_n^A}\sum_n p_n F_\mathrm{I}(\rho_i^B) \nonumber \\
&=\max_{q_n,\boldsymbol{\alpha}_n}\frac{1}{2} \left(1+ \sum_nq_n\left|b_2+\boldsymbol{s}\cdot\boldsymbol{\alpha}_n\right| \right), \label{eq:Fa} 
\end{align}
where the maximization in the last expression is performed over all vectors $\boldsymbol{\alpha}_n$ and probabilities $0\leq q_n\leq1$ such that $\sum_n q_n =1$, $|\boldsymbol{\alpha}_n|\leq 1$ and $\sum_n q_n\boldsymbol{\alpha}_n=0$. 

If $|b_2|\geq |\boldsymbol{s}|$, then using the conditions $|\boldsymbol{\alpha}_n|\leq 1$ and $\sum_n q_n\boldsymbol{\alpha}_n=0$ we immediately obtain
\begin{equation}
 \sum_nq_n\left|b_2+\boldsymbol{s}\cdot\boldsymbol{\alpha}_n\right|=|b_2|\label{eq:FaSum}
\end{equation}
 for any choice of $q_n$ and $\boldsymbol{\alpha}_n$. This directly implies that $F_{\mathrm{a}}(\rho^{AB})=1/2+|b_2|/2$ in this case, in accordance with Eq.~(\ref{eq:FaQubits}).

We now consider the case if $|b_2| < |\boldsymbol{s}|$. We will show that in the maximization in Eq.~(\ref{eq:Fa}) it is enough to consider POVMs consisting of two elements. For a given set of vectors $\boldsymbol{\alpha}_n$ and probabilities $q_n$ we introduce two sets, depending whether $b_2+\boldsymbol{s}\cdot\boldsymbol{\alpha}_n$ is positive or negative:
\begin{subequations}
\begin{align}
S_0&=\{n: b_2+\boldsymbol{s}\cdot\boldsymbol{\alpha}_n\geq0\},\\
S_1&=\{j: b_2+\boldsymbol{s}\cdot\boldsymbol{\alpha}_j<0\}.
\end{align}
\end{subequations}
Using these sets, we express the sum $ \sum_nq_n\left|b_2+\boldsymbol{s}\cdot\boldsymbol{\alpha}_n\right|$ as follows:
\begin{align}
\sum_{n}q_{n}\left|b_{2}+\boldsymbol{s}\cdot\boldsymbol{\alpha}_{n}\right| & =\left(\sum_{n\in S_{0}}q_{n}\right)\left|b_{2}+\frac{\sum_{n\in S_{0}}q_{n}(\boldsymbol{s}\cdot\boldsymbol{\alpha_n})}{\sum_{n\in S_{0}}q_{n}}\right|\\
 & +\left(\sum_{j\in S_{1}}q_{j}\right)\left|b_{2}+\frac{\sum_{j\in S_{1}}q_{j}(\boldsymbol{s}\cdot\boldsymbol{\alpha_j})}{\sum_{j\in S_{1}}q_{j}}\right|.\nonumber 
\end{align}
In the next step, we introduce the probabilities $\Tilde{q}_0=\sum_{n\in S_0}q_n$, $\Tilde{q}_1=\sum_{j\in S_1}q_j$ and vectors 
\begin{subequations}
\begin{align}
\tilde{\boldsymbol{\alpha}}_{0} & =\frac{\sum_{n\in S_{0}}q_{n}\boldsymbol{\alpha}_{n}}{\sum_{n\in S_{0}}q_{n}},\\
\tilde{\boldsymbol{\alpha}}_{1} & =\frac{\sum_{j\in S_{1}}q_{j}\boldsymbol{\alpha}_{j}}{\sum_{j\in S_{j}}q_{j}}.
\end{align}
\end{subequations}
Noting that 
\begin{subequations}
\begin{align}
b_{2}+\boldsymbol{s}\cdot\tilde{\boldsymbol{\alpha}}_{0}&\geq 0, \\
b_{2}+\boldsymbol{s}\cdot\tilde{\boldsymbol{\alpha}}_{1} &< 0,
\end{align}
\end{subequations}
we further obtain the following result: 
\begin{align}
    \sum_{n}q_{n}\left|b_{2}+\boldsymbol{s}\cdot\boldsymbol{\alpha}_{n}\right|&=\tilde{q}_{0}|b_{2}+\boldsymbol{s}\cdot\tilde{\boldsymbol{\alpha}}_{0}|+\tilde{q}_{1}|b_{2}+\boldsymbol{s}\cdot\tilde{\boldsymbol{\alpha}}_{1}| \nonumber \\
    &=\tilde{q}_{0}(b_{2}+\boldsymbol{s}\cdot\tilde{\boldsymbol{\alpha}}_{0})-\tilde{q}_{1}(b_{2}+\boldsymbol{s}\cdot\tilde{\boldsymbol{\alpha}}_{1}).
\end{align}
The vectors $\tilde{\boldsymbol{\alpha}}_n$ and probabilities $\tilde{q}_n$ fulfill the conditions $\sum_n \tilde{q}_n=1$, $|\tilde{\boldsymbol{\alpha}}_n|\leq 1$, and $\sum_n \tilde{q}_n \tilde{\boldsymbol{\alpha}}_n=0$. This implies that they correspond to a two-element POVM on Alice's side via the relation in Eq.~(\ref{eq:QubitPOVM}). 

The arguments just presented show that the maximum in Eq.~(\ref{eq:Fa}) can be achieved with two vectors $\boldsymbol{\alpha}_0$ and $\boldsymbol{\alpha}_1$ and two probabilities $q_0$ and $q_1$ having the properties $0\leq q_0\leq 1$, $q_1 = 1-q_0$,  $|\boldsymbol{\alpha}_n|\leq 1$, $\sum_iq_n \boldsymbol{\alpha}_n=0$. To complete the proof, we will show that the optimal solution is obtained for 
\begin{subequations} \label{eq:solution}
\begin{align}
q_{0} & =q_{1}=\frac{1}{2},\\
\boldsymbol{\alpha}_{0} & =-\boldsymbol{\alpha}_{1}=\frac{\boldsymbol{s}}{|\boldsymbol{s}|}.
\end{align}
\end{subequations}
Recalling that $|b_2|\leq |\boldsymbol{s}|$, the values in Eq.~(\ref{eq:solution}) immediately give a lower bound on the assisted fidelity of imaginarity:
\begin{equation}
F_{\mathrm{a}}(\rho^{AB})\geq \frac{1}{2}(1+|\boldsymbol{s}|). \label{eq:FaBound}
\end{equation}
Let now $q_n$ and $\boldsymbol{\alpha}_n$ be optimal probabilities and vectors [not necessarily coinciding with Eq.~(\ref{eq:solution})]. Without loss of generality we can assume that\footnote{Otherwise, if $b_{2}+\boldsymbol{s}\cdot\boldsymbol{\alpha}_{n}$ is positive (or negative) for all $n$, we obtain $\sum_{n}q_{n}\left|b_{2}+\boldsymbol{s}\cdot\boldsymbol{\alpha}_{n}\right|=|b_2|$. Since $|b_2| < |\boldsymbol{s}|$, this means that we will not be able to reach the maximal value.}
\begin{subequations}
\begin{align}
b_{2}+\boldsymbol{s}\cdot\boldsymbol{\alpha}_{0} & \geq0,\\
b_{2}+\boldsymbol{s}\cdot\boldsymbol{\alpha}_{1} & <0.
\end{align}
\end{subequations}
For the assisted fidelity of imaginarity we thus obtain
\begin{equation}
F_{\mathrm{a}}(\rho^{AB})=\frac{1}{2}[q_{0}(b_{2}+\boldsymbol{s}\cdot\boldsymbol{\alpha}_{0})-q_{1}(b_{2}+\boldsymbol{s}\cdot\boldsymbol{\alpha}_{1})]+\frac{1}{2}. \label{eq:Fa3}
\end{equation}
Since $q_0+q_1=1$, it must be that either $q_0\leq 1/2$ or $q_1\leq 1/2$. In the first case we rewrite Eq.~(\ref{eq:Fa3}) as follows:
\begin{equation}
F_{\mathrm{a}}(\rho^{AB}) = \frac{1-b_2}{2}+q_{0}(b_{2}+\boldsymbol{s}\cdot\boldsymbol{\alpha}_{0}) \leq \frac{1}{2}(1+|\boldsymbol{s}|).
\end{equation}
In the second case ($q_1\leq 1/2$), we rewrite Eq.~(\ref{eq:Fa3}) as
\begin{equation}
    F_{\mathrm{a}}(\rho^{AB})=\frac{1+b_{2}}{2}-q_{1}(b_{2}+\boldsymbol{s}\cdot\boldsymbol{\alpha}_{1})\leq \frac{1}{2}(1+|\boldsymbol{s}|).
\end{equation}
Thus, for $|b_2|<|\boldsymbol{s}|$ the assisted fidelity of imaginarity is bounded above as
\begin{equation}
    F_{\mathrm{a}}(\rho^{AB}) \leq \frac{1}{2}(1+|\boldsymbol{s}|).
\end{equation}
Together with Eq.~(\ref{eq:FaBound}) this proves that $F_{\mathrm{a}}(\rho^{AB})=1/2+|\boldsymbol{s}|/2$ in this case, and the proof of the theorem is complete.

Theorem~\ref{thm:FaQubit} has few surprising consequences. If a two-qubit state has the property $|b_2|\geq |\boldsymbol{s}|$, then the assisted fidelity of imaginarity coincides with the fidelity of imaginarity of Bob's local state: $F_{\mathrm{a}}(\rho^{AB})=(1+ |b_2|)/2$. Thus, in this case Bob will not gain any advantage from assistance, as he can obtain the maximal fidelity by performing a local real operation without any communication. 
For example, let us consider a quantum state shared by Alice and Bob
\begin{equation}
    \rho^{AB}=\frac{p}{2}\openone^A\otimes\ketbra{\hat{+}}{\hat{+}}+(1-p)\ketbra{\phi^+}{\phi^+}
\end{equation}
where we have $b_2=p$ and $s=(0,p-1,0)$. Then if $p=1$, then $\rho^{AB}$ is a product pure state, then no matter what Alice does, Bob can always get the maximal imaginary state $\ket{\hat{+}}$. 
If $\frac{1}{2}<p<1$, the state $\rho^{AB}$ has nonzero entanglement, but we have $|b_2|>|s|$. If Alice chooses a projective measurement along $\boldsymbol{\alpha}$, then Bob will get states with Bloch vector $\boldsymbol{b}\pm E^T\cdot\boldsymbol{\alpha}$ with equal probability. Then the average fidelity with maximally imaginary state reads $\frac{1}{2}\left(|p+(1-p)\alpha_2|+|p-(1-p)\alpha_2|\right)$. As we have $\frac{1}{2}<p<1$, $|(1-p)\alpha_2|<p$, then the average fidelity reads $p$. 
For all other two-qubit states the proof of Theorem~\ref{thm:FaQubit} provides an optimal procedure for obtaining maximal fidelity of imaginarity on Bob's side. For this, Alice needs to perform a von Neumann measurement in the basis $\{\ket{\psi_0},\ket{\psi_1}\}$, where $\ket{\psi_0}$ has the Bloch vector $\boldsymbol{s}/|\boldsymbol{s}|$. The outcome of the measurement is communicated to Bob, who leaves his state untouched if the outcome was $0$, and otherwise applies the real unitary $i\sigma_2$. \question{Needs to be checked}

\section{Proof of Lemma 1}
Note that, the geometric measure of imaginarity and the concurrence of imaginarity are given by \cite{Uhlmann_2000,varun_im}
\begin{eqnarray}\label{geom}
   G(\rho)=\min_e \sum_jp_{j}\frac{1-|\braket{\psi_j^*}{\psi_j}|}{2}=\frac{1-\sqrt{F(\rho,\rho^T)}}{2}
\end{eqnarray}
\begin{eqnarray}\label{conc}
    C(\rho)=\min_e\sum p_j |\braket{\psi_j^*}{\psi_j}|=\max\left\{0,\lambda_1-\sum_{k>1}\lambda_k\right\}
\end{eqnarray}
In the above $\max_e$ and $\min_e$ are maximisation and minimisation over pure state ensembles of $\rho$. Whereas, $\{\lambda_1,\lambda_2...\}$ are the eigenvalues (in decreasing order) of $(\sqrt{\rho}\rho^T\sqrt{\rho})^{\frac{1}{2}}$. In general, for probabilistic transformations, the following inequality holds
\begin{equation}\label{prob.bound}
    p(\rho\rightarrow\sigma)\leq\min\left\{\frac{G(\rho)}{G(\sigma)},1\right\}.
\end{equation}
It was further shown in \cite{varun_im}, that the optimal probability of converting a pure state $\psi$ to a arbitrary quantum state $\rho$ is given by
\begin{equation}\label{prob.eq}
    p(\psi\rightarrow \rho)=\min\left\{\frac{G(\psi)}{G(\rho)},1\right\}.
\end{equation}

In a one way LQRCC procedure, Alice performs a general quantum measurement and corresponding to the outcomes (with probabilites $\{p_j\}$) of Alice, Bob's local state is found in the state $\rho_j$, such that, $\{p_j,\rho_i\}$ is an ensemble of $\rho^B$. Conditioned on the outcome of Alice ($i$), Bob can perform a local stochastic real operation on $\rho_i$, probabilistically converting it into $\sigma^B$. Using Eq. (\ref{prob.bound}) and Eq. (\ref{prob.eq}), it follows
\begin{equation}\label{assisted}
    P_a\leq\sum_{j} p_j\min\left\{\frac{G(\rho_j)}{G(\sigma_{B})},1\right\}\leq \sum_{jk}p_jq_k\min\left\{\frac{G(\psi_{j,k})}{G(\sigma_{B})},1\right\}.
\end{equation}
The second inequality follows from Eq.(\ref{geom}), $G(\rho_j)$ is calculated by minimising over all pure state ensembles of $\rho_j$. Therefore, the second inequality holds for any pure state decomposition of $\rho_j$, like $\{q_k,\psi_{jk}\}$. Note that $\{p_jq_k, \psi_{jk}\}$ is a pure state decomposition of $\rho^B$. Note that, any pure state decomposition of $\rho^B$ can be realised by a suitable local measurement by Alice. Using this fact, along with Eq.(\ref{geom}) and Eq.(\ref{prob.eq}) implies that
\begin{align*}
P_a &= \min\left\{\frac{1-\min_e\sum_kp_k|\braket{\psi_k}{\psi_k^{*}}|}{2G(\sigma^{B})},1\right\}\\&=\min\left\{\frac{1-C(\rho^B)}{1-\sqrt{F(\sigma^B,(\sigma^B)^T)}},1\right\}.
\end{align*}
Here, $\min_e$ is the minimisation over pure state ensembles of $\rho^B$. This completes the proof.

\subsection{Proof of Theorem \ref{thm:PureConversion}}
From Lemma 1, we know that optimal probability for Bob to locally achieve $\sigma^B$ from a shared bipartite pure state $\psi^{AB}$ with unit fidelity, via LQRCC is given by
\begin{equation}
    P(\psi^{AB} \rightarrow \sigma^{B}) =\min \left\{\frac{1-\mathscr{I}_c(\rho^B)}{2\mathscr{I}_g(\rho)},1\right\}. 
\end{equation}
If we want to achieve $\sigma^B$ with fidelity at least $f$, the best strategy is to go to a state ($\sigma'^B$), within the fidelity ball around $\sigma^B$, with a minimal geometric measure of imaginarity. Therefore,
\begin{equation}
P_{f}(\psi^{AB}\rightarrow\sigma^B)=\min\left\{\frac{1-\mathscr{I}_c(\rho^B)}{2\mathscr{I}_g(\sigma'^B)},1\right\}.
\end{equation}
From \cite{varun_im}, we know that
\begin{align}
\mathscr{I}_g(\sigma'^B) & =\sin^{2}\left(\max\left\{ \sin^{-1}\!\sqrt{\mathscr{I}_g(\sigma^B)}-\cos^{-1}\!\sqrt{f},0\right\} \right).
\end{align}
We now define
\begin{equation}
    m=\sin^{-1}\sqrt{\frac{1-\mathscr{I}_c(\rho^B)}{2}}-\sin^{-1}\sqrt{\mathscr{I}_g(\sigma^B)} +\cos^{-1}\sqrt{f}.
\end{equation}
First, consider the case when $m\geq0$, which implies
\begin{equation}
    \sin^{-1}\sqrt{\mathscr{I}_g(\sigma^B)}-\cos^{-1}\sqrt{f} \leq \sin^{-1}\sqrt{\frac{1-\mathscr{I}_c(\rho^B)}{2}}.
\end{equation}
We know that
\begin{equation}
\sin^{-1}\sqrt{\mathscr{I}_g(\sigma^B)}-\cos^{-1}\sqrt{f}\in[-\pi/2,\pi/4]
\end{equation}
and $\sin^{-1}\sqrt{\frac{1-\mathscr{I}_c(\rho^B)}{2}}\in[0,\pi/4]$. Therefore, 
\begin{equation}
    \max\left\{\sin^{-1}\sqrt{\mathscr{I}_g(\sigma^B)}-\cos^{-1}\sqrt{f},0\right\}\leq \sin^{-1}\sqrt{\frac{1-\mathscr{I}_c(\rho^B)}{2}}.
\end{equation}
Using these results, we get
\begin{align}
\mathscr{I}_g(\sigma^{B'}) & =\sin^{2}\left(\max\left\{ \sin^{-1}\sqrt{\mathscr{I}_g(\sigma^B)}-\cos^{-1}\sqrt{f},0\right\} \right)\\
 & \leq\sin^{2}\left(\sin^{-1}\sqrt{\frac{1-\mathscr{I}_c(\rho^B)}{2}}\right)=\frac{1-\mathscr{I}_c(\rho^B)}{2}. \nonumber
\end{align}
For the case when $\frac{1-\mathscr{I}_c(\rho^B)}{2} > 0$, the above inequality implies
\begin{equation}
    \frac{1-\mathscr{I}_{c}(\rho^B)}{2\mathscr{I}_{g}(\sigma^{B'})} \geq 1.
\end{equation}
This shows that $P_{f}(\psi^{AB}\rightarrow\sigma^B)=1$ when $m\geq 0 $. 

Now, we look at the other case when $m<0$, i.e.,
\begin{align}
   \sin^{-1}\sqrt{\mathscr{I}_g(\sigma^B)}-\cos^{-1}\sqrt{f}>\sin^{-1}\sqrt{\frac{1-\mathscr{I}_c(\rho^B)}{2}}>0.
\end{align}
From the above inequality and Lemma 1, we have
\begin{equation}\label{eq:optimal_probability_imaginarity2}
    P_{f}(\psi^{AB}\rightarrow\sigma^B)=\frac{1-\mathscr{I}_c(\rho^B)}{2\sin^{2}(\sin^{-1}\sqrt{\mathscr{I}_g(\sigma^B)}-\cos^{-1}\sqrt{f}) }.
\end{equation}
Using the above result, a closed expression can also be found for $F_p$. Let's first consider the case when $p \leq \frac{1-\mathscr{I}_c(\psi^{AB})}{2\mathscr{I}_g(\sigma^B)}<1$, in this case $F_p (\psi \rightarrow \sigma^B) = 1$ (follows from Lemma 1). When $1\geq p>\frac{\mathscr{I}_g(\psi)}{\mathscr{I}_g(\sigma^B)}$, the optimal achievable fidelity can be obtained by solving Eq.~\eqref{eq:optimal_probability_imaginarity} for $f$, which gives 
\begin{eqnarray}
F_p (\psi^{AB} \rightarrow \sigma^B)=\cos^{2}\left[\sin^{-1}\!\sqrt{\mathscr{I}_g(\sigma^B)}-\sin^{-1}\!\sqrt{\frac{1-\mathscr{I}_c(\rho^B)}{2p}}\right].\nonumber
\end{eqnarray}
This completes the proof.

\section{SDP upperbounds for state transformations}
As we already mentioned, for any real CP map $\Lambda:R\rightarrow R'$, $\Gamma^{\Lambda}_{RR'}$ is the corresponding choi matrix of $\Gamma^\Lambda_{RR'}$, given by
\begin{equation}
   \Gamma^\Lambda_{RR'}=\openone\otimes\Lambda\left(\sum_{j,k}\ket{j}\!\bra{k}\otimes\ket{j}\!\bra{k}\right).
\end{equation}
 It follows that (see Eq. (4.2.12) of \cite{https://doi.org/10.48550/arxiv.2011.04672}),
\begin{equation}\label{choi}
    \Lambda(\rho_R)=\tr_{R}(\Gamma^\Lambda_{RR'}(\rho_R^{T}\otimes \openone_{R'})).
\end{equation}
For any pure state $\ket{\psi_{R'}}$ 
\begin{eqnarray}\label{inner}
    \bra{\psi_{R'}}\Lambda(\rho_R)\ket{\psi_{R'}}&=\tr(\Gamma^\Lambda_{RR'}(\rho_R^{T}\otimes \ket{\psi_{R'}}\!\bra{\psi_{R'}})).
\end{eqnarray}
Using the fact that, choi matrices of LQRCC operations are invariant under partial transpose, one can give a SDP computable upperbound for the optimal achievable fidelity for a given probability $F_p(\rho_{AB}\rightarrow\ket{\psi_{AB}})$: \\
Maximise:
\begin{equation}
  \frac{1}{p} \tr(X_{ABA'B'}\rho_{AB}^{T}\otimes \ket{\psi_{A'B'}}\!\bra{\psi_{A'B'}})\,\, 
\end{equation}
under the constraints,
\begin{eqnarray}\nonumber
   && X_{ABA'B'}\geq 0,\, 
     X_{ABA'B'}^{T_{BB'}}=X_{ABA'B'},  \tr_{A'B'} X_{ABA'B'}\leq\id_{AB}\,\,\textrm{and}\,\,\\
     &&\tr(X_{ABA'B'}\rho_{AB}^{T}\otimes \id_{B'})=p.
\end{eqnarray}


\section{Quantum Chernoff divergence and scaling of asymptotic imaginarity distillation}
 Fidelity of imaginarity $F_{\mathrm{I}}$, quantifies the maximum achievable fidelity between a state $\rho$ and the maximally imaginary state. It can be expressed as
\begin{equation}
    F_\mathrm{I}(\rho)=\max_{\Lambda}F(\Lambda(\rho), \ket{\MI}\bra{\MI}) =\frac{1}{2}+\frac{1}{4}||\rho - \rho^{T}||_{1} .
\end{equation}
Here, the maximisation is performed over all real CPTP maps. If we have $n$ copies of $\rho$, we can write
\begin{equation}
    F_\mathrm{I}(\rho^{\otimes n}) = \frac{1}{2}+\frac{1}{4}||\rho^{\otimes n} - (\rho^{T})^{\otimes n}||_{1}.
\end{equation}
 If $\rho$ is a pure state, i.e., $\rho =\ket{\psi}\bra{\psi}$, then we can calculate fidelity of imaginarity of multiple copies as
\begin{eqnarray}
    F_\mathrm{I}(\ket{\psi}\bra{\psi}^{\otimes n}) &=& \frac{1}{2}+\frac{1}{4}||\ket{\psi}\bra{\psi}^{\otimes n} - (\ket{\psi}\bra{\psi}^{T})^{\otimes n}||_{1} \nonumber\\
    &=& \frac{1}{2}+\frac{1}{4}||\ket{\psi}\bra{\psi}^{\otimes n} - (\ket{\psi^{*}}\bra{\psi^{*}})^{\otimes n}||_{1} \nonumber\\
    &=& \frac{1}{2}+\frac{1}{2}\sqrt{1 -|\braket{\psi^{*}}{\psi}|^{2n}}. 
\end{eqnarray}
For general states, to see the behaviour of $F_{\mathrm{I}}(\rho^{\otimes n})$, with increasing $n$, consider the quantity $P=1-F_{\mathrm{I}}(\rho^{\otimes n})$. From Ref.~\cite{PhysRevLett.98.160501}, it follows that the following limit exists and is equal to the quantum Chernoff divergence between $\rho$ and $\rho^T$:
\begin{eqnarray}\label{chernoff}
 \lim_{n\rightarrow\infty} \frac{-\log P}{n}= \chi(\rho,\rho^{T})=-\log(\min_{0\leq s\leq 1}\tr (\rho^s(\rho^{T})^{1-s})).
\end{eqnarray}
One can analytically perform this minimisation and show that minimum value is attained at $s=1/2$. In order to show this fact, let's assume that the spectral decomposition of $\rho$ is given by 
\begin{equation}
    \rho =\sum_{j}p_j\ket{\psi_j}\bra{\psi_j},
\end{equation}
and therefore 
\begin{equation}
    \rho^T =\sum_{j}p_j\ket{\psi^{*}_j}\bra{\psi^{*}_j}.
\end{equation}
The Chernoff divergence is given by
\begin{eqnarray}
  \chi(\rho,\rho^{T})&=  -\log(\min_{0\leq s\leq 1}\tr(\sum_{j}p^s_j\ket{\psi_j}\bra{\psi_j})(\sum_{k}p^{1-s}_k\ket{\psi^{*}_j}\bra{\psi^{*}_j}))\nonumber\\
  &=-\log(\min_{0\leq s\leq 1}\sum_{j,k}p^s_ip^{1-s}_k|\braket{\psi_j}{\psi^{*}_k}|^2).
\end{eqnarray}
Note that, $|\braket{\psi_j}{\psi^{*}_k}|=|\braket{\psi_k}{\psi^{*}_j}|$. This implies that
\begin{equation}
  \chi(\rho,\rho^{T})=-\log(\min_{0\leq s\leq 1}\sum_{j\leq k}(p^s_jp^{1-s}_k+p^s_kp^{1-s}_j)|\braket{\psi_j}{\psi^{*}_k}|^2)
\end{equation}
here, $p^{s}_jp^{1-s}_k+p^s_kp^{1-s}_j\geq 2\sqrt{p_jp_k}$. This follows from AM-GM inequality, which says $\frac{a+b}{2}\geq\sqrt{ab}$ for all $a,b\geq 0$. This lower bound (minimum value) is attained at $s=1/2$. This proves that 
\begin{eqnarray}
   \chi(\rho,\rho^{T})&= -\log(\tr\sqrt{\rho\rho^{T}}).
\end{eqnarray}
Therefore, from Eq. (\ref{chernoff}), it follows that asymptotically the fidelity of imaginarity behaves as
\begin{eqnarray}
     F_\mathrm{I}(\rho^{\otimes n})&\sim 1-\exp(-n\cdot \chi(\rho,\rho^{T}))\\
     &=1-(\tr\sqrt{\rho\rho^{T}})^n \nonumber
\end{eqnarray}

\section{Proof of the relation between channel discrimination and state discrimination}\label{app:channel discrimination equivalence}
Here we demonstrate a clear link between the task of ancilla-free channel discrimination and the task of LOCC discrimination of bipartite states, the latter studied in Refs.~\cite{Wootters1990,PRLversion,PRAversion}.
Specifically, we consider the following two scenarios:
\begin{enumerate}
    \item Let $\mathcal{N}$ and $\mathcal{M}$ be two real channels from $A$ to $B$, chosen with equal probability $\frac{1}{2}$. If we want to discriminate between them in an ancilla-free scenario better than with a random guess, we must find a real state $\rho$ of $A$ and a real POVM element $E$ of $B$ such that $\tr\left[E\mathcal{N}\left(\rho\right)\right]\neq\tr\left[E\mathcal{M}\left(\rho\right)\right]$. 
    Notice that this protocol does not involve any bipartite input states and bipartite effects.
\item Let $\mathcal{N}$ and $\mathcal{M}$ be two real channels from $A$ to $B$. This time, we bring in the maximally entangled state $\phi^+=\ketbra{\phi^+}{\phi^+}^{AA'}$, between systems $A$ and $A'$ ($A'$ is a copy of $A$), where $\ket{\phi^+}=\sum_j\ket{jj}/\sqrt{d_A}$, and $d_A$ is the dimension of $A$. We apply $\mathcal{N}$ and $\mathcal{M}$ only to the $A'$ part of this maximally entangled state. This  results in two bipartite states between systems $A$ and $B$, $N^{AB}$ and $M^{AB}$, respectively, which are the normalized Choi states of the two channels $\mathcal{N}$ and $\mathcal{M}$. Now consider the task of discriminating between these two bipartite states of $AB$ using \emph{only local} real measurements. Again, if we want to discriminate between them better than with a random guess, we must find a real POVM element $E$ of system $A$ and a real POVM element $F$ of system $B$ such that $\tr\left[\left(E\otimes F\right) N^{AB}\right]\neq\tr\left[\left(E\otimes F\right) M^{AB}\right]$. 
\end{enumerate}

In the following we show that these two scenarios produce the same probabilites when POVMs are applied to states. 
Note that we can reconstruct the action of a channel on a state from its normalized Choi state: if $\mathcal{N}$ is a channel from $A$ to $B$, $\rho$ is a state of $A$, we have that  $\mathcal{N}\left(\rho\right)$ can be written in terms of the normalized Choi state $N^{AB}$ as \begin{equation}\label{eq:Choi}\mathcal{N}\left(\rho^A\right)=d_A\tr_A\left\{\left[\left(\rho^A\right)^{T}\otimes\id^B\right]N^{AB}\right\},\end{equation}where $d_A$ is the dimension of the input system $A$. Thus, if $E$ is a (real) POVM element on $B$, omitting system superscripts for simplicity, we have
\begin{align}
\tr \left[E\mathcal{N}\left(\rho\right)\right]&=d_A\tr\left[\left(\rho^T\otimes E\right)N\right]\nonumber\\
&=\tr\left[\left(\frac{1}{\sqrt{d_A}}\rho^T\otimes \frac{1}{\sqrt{d_A}}E\right)N\right].
\end{align}
Note that $\mathbf{0}\leq\frac{1}{\sqrt{d_A}}\rho^T\leq\id$ and $\mathbf{0}\leq \frac{1}{\sqrt{d_A}}E\leq\id$, then $\frac{1}{\sqrt{d_A}}\rho^T$ and $\frac{1}{\sqrt{d_A}}E$ are both valid (real) POVM elements on $A$ and $B$, respectively. So now we have an LOCC discrimination scenario on the normalized Choi state $N^{AB}$ that yields exactly the same probability as the original ancilla-free channel discrimination scenario. 

Conversely, let us consider the LOCC discrimination scenario of normalized  Choi states. Let $N^{AB}$ be the normalized Choi state of a channel $\mathcal{N}$ from $A$ to $B$. If $E$ and $F$ are POVM elements on $A$ and $B$, respectively, we want to calculate the probability $\tr\left[\left(E\otimes F\right)N^{AB}\right]$. Note that, assuming $E\neq \mathbf{0}$, $\rho:=\frac{1}{\tr E}E$ is a valid quantum state, so $\tr\left[\left(E\otimes F\right)N^{AB}\right]=\tr E \tr\left[\left(\rho\otimes F\right)N^{AB}\right]$. Then, we have
\begin{align}
    \tr E \tr_{AB}\left[\left(\rho\otimes F\right)N^{AB}\right]&=\tr E\tr_{B}\left\{F\tr_A\left[\left(\rho\otimes \id\right)N^{AB}\right]\right\}\nonumber\\
    &=\frac{\tr E}{d_A} \tr_B\left[F\mathcal{N}\left(\rho^T\right)\right]\nonumber\\
    &=\tr\left[F'\mathcal{N}\left(\rho^T\right)\right],
\end{align}
where we have used Eq.~\eqref{eq:Choi}, and we have defined $F':=\frac{\tr E}{d_A} F$. Now, $\rho^T$ is still a valid quantum state of $A$, and $F'$ is still a valid POVM element on $B$ because $\frac{\tr E}{d_A}\leq 1$. So now we have an ancilla-free discrimination scenario on the channels associated with the bipartite normalized state that yields exactly the same
probability as the original bipartite LOCC discrimination scenario. In this way, we have proven that all probabilities arising in one of the two scenarios can be completely reproduced by the other scenario, so they are in some sense equivalent in terms of the probabilities they can generate.

Having established the relation of channel discrimination and local discrimination of their corresponding Choi states, we can see that the advantage of imaginarity in real channel discrimination shows up when both initial probe state and measurement contain imaginarity. We accomplish this by mapping the ancilla-free channel discrimination scenario into the LOCC state discrimination scenario, using (normalized) Choi matrices, as discussed above. Let us consider the example of a qubit channel $\mathcal{N}$. Note that its (normalized) Choi state can be written as 
\begin{align}\label{eq:proindividual}
&N^{AB}\nonumber\\
&=\frac{1}{2}\left(\id+\sum_ja_j\sigma_j^A\otimes\id^B+\id^A\otimes\sum_jb_j\sigma_j^B+\sum_{j,k}E_{jk}\sigma_j^A\otimes\sigma_k^B\right), 
\end{align} where $i,j\in \left\{x,y,z\right\}$, and the $\sigma_j$'s are Pauli matrices. If $\mathcal{N}$ is a real operation, then we can conclude that the only term containing $\sigma_y$ must only be $\sigma_y\otimes\sigma_y$. Recall that $\tr\left[S \sigma_y\right]=0$ for any real symmetric $2\times 2$ matrix $S$ (cf.\ Ref.~\cite{PRLversion}). For this reason, any   POVM element $M^{AB} = E^A \otimes F^B$, with real symmetric matrices $E$ or $F$, cannot be used to detect the presence of the $\sigma_y\otimes\sigma_y$ term in a Choi matrix of a real operation. Consequently, there are some real operations that are perfectly distinguishable, but become indistinguishable using  an ancilla-free protocol if we only use real states and measurements. However, if we are still restricted to real probe states and measurements, but we allow an ancilla, then the same real operations become perfectly distinguishable again. To understand why, notice that when we allow an ancilla, we can use the state  $\phi^+$ as probe state for all real operations, thus producing their normalized Choi states. Then the task becomes distinguishing between their Choi states, but \emph{without any LOCC constraints} (recall that the LOCC constraint comes from the ancilla-free scenario). Removing the LOCC constraint from the discrimination of the Choi states makes the advantage provided by imaginarity disappear. Consequently, with an ancilla, we can perform as well with just real states and measurements as we do with non-real ones.

\section{Experimental details}
In Module \textbf{A}, two type-I phase-matched $\beta$-barium borate (BBO) crystals, whose optical axes are normal to each other, are pumped by a continuous laser at 404 nm, with a power of 80 mW, for the generation of photon pairs with a central wavelength at $\lambda=808$ nm via a spontaneous parametric down-conversion process (SPDC). A half-wave plate (HWP) and a quarter-wave plate (QWP) working at 404 nm set before the lens and BBO crystals is used to control the polarization of the pump laser. Two polarization-entangled photons are generated and then distributed through two single-mode fibers (SMF), where one represents Bob and the other Alice. Two interference filters (IF) with a 3 nm full width at half maximum (FWHM) are placed to filter out proper transmission peaks. HWPs at both ends of the SMFs are used to control the polarization of both photons. 

In Module \textbf{B} for preparing Werner states, two 50/50 beam splitters (BSs) are inserted into one branch. In the transmission path, the two-photon state is still a Bell state. In the reflected path, three 400$\lambda$ quartz crystals and a HWP with angles set to $22.5^\circ$  are used to dephase the two-photon state into a completely mixed-state $\openone^{AB}/4$. The ratio of the two states mixed at the output port of the second BS can be changed by the two adjustable apertures (AA) for the generation of Werner states. This setup also allows us to implement a class of quantum channels which are specified in the main text.

\end{document}